\documentclass[twocolumn,prb,aps]{revtex4}

\usepackage{graphicx}
\usepackage{dcolumn}
\usepackage{amsmath}
\usepackage{latexsym}

\begin{document}

\title{Noise-enabled precision measurements of a Duffing nanomechanical resonator}

\author{J. S. Aldridge}
\author{A. N. Cleland}
\affiliation{Department of Physics, University of California at Santa Barbara, Santa Barbara, CA
93106}

\date{\today}

\begin{abstract}
We report quantitative experimental measurements of the nonlinear response of a radiofrequency
mechanical resonator, with very high quality factor, driven by a large swept-frequency force. We
directly measure the noise-free transition dynamics between the two basins of attraction that
appear in the nonlinear regime, and find good agreement with those predicted by the one-dimensional
Duffing equation of motion. We then measure the response of the transition rates to controlled
levels of white noise, and extract the activation energy from each basin. The measurements of the
noise-induced transitions allow us to obtain precise values for the critical frequencies, the
natural resonance frequency, and the cubic nonlinear parameter in the Duffing oscillator, with
direct applications to high sensitivity parametric sensors based on these resonators.
\end{abstract}

\maketitle

Doubly-clamped mechanical resonators have recently been the subject of much attention, due to the
ability to make very high frequency, high quality factor resonators, with applications in weak
force and small mass detection, frequency stabilization, and possibly quantum computation
\cite{cleland:2653, Ekinci:2004b, Ilic:2000, Gr94, carr:920, cleland:2758, ClelandandKnobel,
armour:148301, ekinci:2004, Wybourne:2001, turner:1998, turner:2002,Soskin:2003}. The limit for
parametric sensing is often set by the precision with which a resonator parameter, such as the
mass, can be monitored, limited typically by measurement and intrinsic noise sources. Here we show
how one can use the intrinsic nonlinear response of these resonators, and the addition of external
broadband noise, to significantly improve the measurement precision of two such parameters, the
resonance frequency and the cubic nonlinearity. This has direct implications for the ultimate
sensitivity of such parametric sensors.

At large drive amplitudes, doubly-clamped resonators exhibit a bistable response quantitatively
similar to that of the Duffing oscillator \cite{yurkeandgreywall, Nayfeh}. The motion in the
fundamental mode of a doubly-clamped beam is thus well-approximated by the Duffing equation, which
for a natural resonance frequency $\Omega_0$ and quality factor $Q$, driven at frequency $\Omega$,
has the form
\begin{equation}
M\frac{ d^2 Y}{d t^2}+M\frac{\Omega_{0}}{Q}\frac{d Y}{d
t}+M\Omega_{0}^2 Y+K Y^3= {\mathcal B}\cos(\Omega t)+{\mathcal
B}_{noise}(t), \label{eq:duffing}
\end{equation}
where $Y$ denotes the displacement amplitude of the midpoint of the beam, $M$ denotes the mass of
the beam, ${\mathcal B}$ the amplitude of the external driving force, and ${\mathcal B}_{noise}(t)$
the stochastic forcing function due to thermal and external noise \cite{cleland:2758,
yurkeandgreywall, Nayfeh}. This equation assumes that the beam oscillates in the mode with natural
frequency $\Omega_0$, that the displacement amplitude $Y(t)$ is the only relevant degree of
freedom, and that the equation of motion includes only the third-order nonlinearity, with strength
$K$.

The displacement $Y(t)$ in Eq. (\ref{eq:duffing}) can be written as
\begin{equation}
 Y(t) = U_{1}(t) \cos(\Omega t)+U_{2}(t) \sin(\Omega t),
\label{eq:capIandcapQ}
\end{equation}
in terms of the two quadrature amplitudes $U_{1,2}(t)$. For a high $Q$ system driven at frequency
$\Omega$ near $\Omega_0$, the slowly-varying envelope approximation can be used
\cite{yurkeandgreywall,Dykman79}, where the functions $U_{1,2}(t)$ are replaced by their slowly
varying averages, $u_{1,2}(t)$, respectively.

In the absence of noise, the average functions $u_{1,2}(t)$ satisfy the equations of motion
\begin{equation}\label{eq.trajeq}
\left . \begin{array}{rcl}
 \displaystyle \frac{d^2 u_{1}}{d t^2} &=& \displaystyle
 (\Omega^2-\Omega_{0}^2)u_{1}-\frac{3}{4} \frac{K}{M} u_{1}(u_{1}^2+u_{2}^2) \\
 &-& \displaystyle  \frac{\Omega_0}{Q} \Omega u_{2}- \frac{\Omega_0}{Q} \frac{d u_{1}}{d t}
 - 2 \Omega \frac{d u_{2}}{d t} +
 \frac{{\mathcal B}}{M}, \\
 \displaystyle \frac{d^2 u_{2}}{d t^2} &=& \displaystyle (\Omega^2-\Omega_{0}^2)u_{2}-
  \frac{3}{4} \frac{K}{M} u_{2}(u_{1}^2+u_{2}^2) \\
 &+& \displaystyle \frac{\Omega_0}{Q} \Omega u_{1}- \frac{\Omega_0}{Q} \frac{d u_{2}}{d t} +
 2 \Omega \frac{d u_{1}}{d t}.
\end{array} \right \rbrace \quad
\end{equation}

The Duffing oscillator exhibits one stable state for small drive amplitudes ${\mathcal B}$, while
above a critical amplitude ${\mathcal B}_c$ a bifurcation occurs, creating two stable basins of
attraction. One basin corresponds to larger displacement amplitudes, and is stable for drive
frequencies up to an upper critical frequency $\nu_U$ ($\nu = \Omega/2 \pi$), determined by the
drive amplitude ${\mathcal B}$. The other stable basin has smaller displacement amplitude, and is
stable for frequencies down to a lower critical frequency $\nu_L$, also determined by the drive
amplitude. The stable attractors are found by setting all time derivatives in Eq. (\ref{eq.trajeq})
to zero and solving for $u_{1,2}$, yielding three equilibrium points. Two of these equilibrium
points are stable foci, and the third is a metastable separatrix.

A transition between the two basins occurs in the absence of noise when the energy barrier
separating them is reduced to zero, by changing either the drive amplitude or the drive frequency.
In the presence of noise, however, the Duffing oscillator will exhibit stochastic transitions
between the two basins. For weak noise signals, the transitions occur only near the critical
frequencies $\nu_{L,U}$, while as the noise power is increased, the separation between the upper
and lower transition frequencies is effectively reduced.

\begin{figure}
\includegraphics[width=1\linewidth]{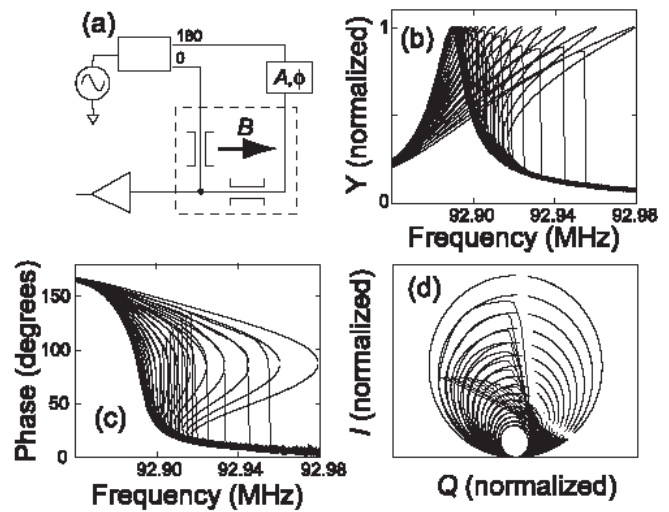} \caption{(a) Schematic showing
circuit and active and reference beams; dotted outline encloses cryogenic part of experiment. Box
labelled $0, 180$ is a $180^{\circ}$ phase splitter, and that labelled $A, \phi$ allows adjustment
of the amplitude and phase of the signal. The arrow indicates the orientation of the magnetic
field. (b) Hysteresis in amplitude versus drive frequency, for drive amplitudes from -68 to -53
dBm, in 1 dBm steps. (c) Hysteresis in phase for the same drive amplitudes as (b). (d) Hysteresis
in $u_1-u_2$ plane, plotted as $I$ versus $Q$ in dimensionless units. } \label{fig:figure1}
\end{figure}

Here we make detailed measurements of the nonlinear dynamics of a doubly-clamped beam,
investigating both the dynamical motion and the change in the inter-basin transition rates due to
broadband noise. Our experimental system comprises a pair of doubly-clamped beams of single-crystal
aluminum nitride, with dimensions $3\times 0.2 \times 0.14~\mu$m$^3$, oriented perpendicular to one
another and fabricated together on a chip of single-crystal Si. The fabrication technique is
described elsewhere \cite{cleland:2070}. The chip was placed in the vacuum bore of an $B = $ 8 T
magnet at 4.2 K, with one beam (the active beam) oriented perpendicular to the field direction, the
other (reference) beam parallel to the field. Magnetomotive actuation and displacement detection
was used to drive the active beam \cite{cleland:2653}, where the parallel orientation of the
reference beam decouples it from the drive force (see Fig. \ref{fig:figure1}). The active beam had
a natural resonance frequency $\nu_0 = \Omega_0/2 \pi = 92.9$ MHz, a quality factor $Q = 6750$, and
a critical drive power for inducing the hysteretic bifurcation of -61 dBm. Using the beam
resistance of 11 $\Omega$, this corresponds to a critical drive force $\mathcal{B}_c =$ 580 pN, and
a midpoint displacement of 18 nm.

Measurements were made with a radiofrequency bridge \cite{ekinci:2253}, as shown in Fig.
\ref{fig:figure1}(a). The rf drive signal is split by a 180$^{\circ}$ phase splitter, with the two
phases passed through separate stainless coaxial cables of similar construction. The 180$^{\circ}$
phase-shifted signal is connected to one end of the reference beam, and the 0$^{\circ}$ signal
connected to one end of the active beam. The other ends of the two beams are connected to a third
coaxial cable that returns to room-temperature electronics. The bridge can be balanced in both
amplitude and phase over the range of frequencies used in this experiment, and is typically tuned
so that the electrical signal is proportional to the displacement-induced electromotive force
\cite{cleland:2653}.

The signal measured in the experiment is the demodulated output of the bridge, giving the in-phase
and out-of-phase quadrature signals $I(t$) and $Q(t)$. These are proportional, to within a phase
factor, to the average amplitudes $u_{1,2}(t)$ \cite{ekinci:2253, cleland:2653}. In Fig.
\ref{fig:figure1}(b)-(d) we display the response of the active beam to a range of drive amplitudes,
where the frequency is swept through the resonance for each drive amplitude; the hysteresis in the
amplitude and phase response is in quantitative agreement with the response expected for a Duffing
oscillator.

\begin{figure}
\includegraphics[width=\linewidth]{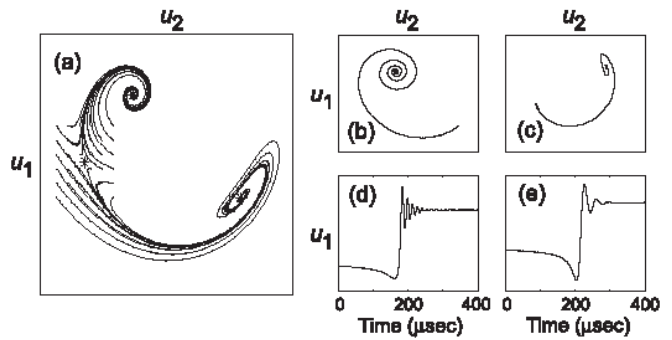}
\caption{(a) Numerically-generated phase-space flow for a drive force 9 dB above the critical point
${\mathcal B}_c$, and drive frequency 40 kHz above $\Omega_0/2 \pi$. Flow begins near the
separatrix and evolves toward either focus. (b) Experimental phase space mean trajectory from focus
1 to focus 2 (8000 averages). (c) Data for phase space mean trajectory from focus 2 to focus 1
(8000 averages). (d) and (e) Experimental time traces for the two switching transitions (8000
averages).} \label{fig:trajectories}
\end{figure}

In Fig. \ref{fig:trajectories} we compare the measured quadrature amplitudes to numerical solutions
of Eq. (\ref{eq.trajeq}). In Fig. \ref{fig:trajectories}(a) we show the calculated phase-space
trajectories, and in (b) and (c) the experimentally measured trajectories. In Fig.
\ref{fig:trajectories}(d) and (e) the time traces are shown for the switching transitions. The
correspondence between the image in (a) and those measured in (b) and (c) is quite clear. It can be
shown that as the drive frequency is varied, the stable points will follow a circle on the
$u_{1,2}$ plane, even in the presence of the Duffing nonlinearity. These circles are evident in
Fig. \ref{fig:figure1}(d).

We now turn to a discussion of the noise-induced transitions between the stable foci. The problem
of thermally-activated escape from a potential landscape with a single basin of attraction is a
thoroughly studied problem \cite{Kramers}. The escape rate over a barrier of height $E_B$ is given
by $\Gamma = a(Q)\nu_0 \exp(-E_{B}/k_B T)$, determined predominantly by the Arrhenius factor and
less so by the $Q$-dependent prefactor $a(Q)$. Our system differs from this classic problem: Here,
there is a basin of attraction about each of the two foci found on a Poincar\'e map of the
configuration space. Instead of a one-dimensional potential well, there is a quasipotential, with
the dynamics governed by the noise energy at each point in the configuration space
\cite{Kautz87_pla}. The equivalent activation energy, $E_{A}$, for transitions between the foci, is
found by integrating the minimum available noise energy over the trajectory between the foci.

Transitions were induced by using an external broadband white noise signal, combined with the
radiofrequency drive signal using a directional coupler, to generate a signal that included both
the drive signal ${\mathcal B}$ and the noise signal ${\mathcal B}_{n}$. Typical noise powers
ranged from -130 to -100 dBm/Hz. The drive signal itself was produced by a source with very low
phase noise; with no additional noise power, transitions were still induced by this remnant phase
noise, to which the resonator is very sensitive. The thermal noise of the circuit, and the
mechanical noise associated with the finite resonator $Q$, are estimated to be 70 dB below the
source phase noise, and were too small to induce measurable transitions in the system.

Transition histograms were measured by applying a drive signal to the resonator above the critical
value, preparing the resonator in one of the two basins of attraction, and monitoring the switching
transitions to the other basin. We measured histograms of the switching probability per unit time,
$h(t)$, by sweeping the drive frequency $\nu(t) = \Omega(t)/2 \pi$ at a constant rate $s = d \nu/d
t$, and recording the drive frequency at which a transition occurred. This is a technique that has
been extensively used for measuring switching distributions in current-biased Josephson junctions
\cite{fulton:1974}. The transition rate $\Gamma(\nu)$ is extracted from the histogram $h(t)$ using
$\Gamma(\nu(t))=(1-\int_{-\infty}^{t}h(t') dt')^{-1} \, s \, h(t)$.

\begin{figure}
\includegraphics[width=1\linewidth]{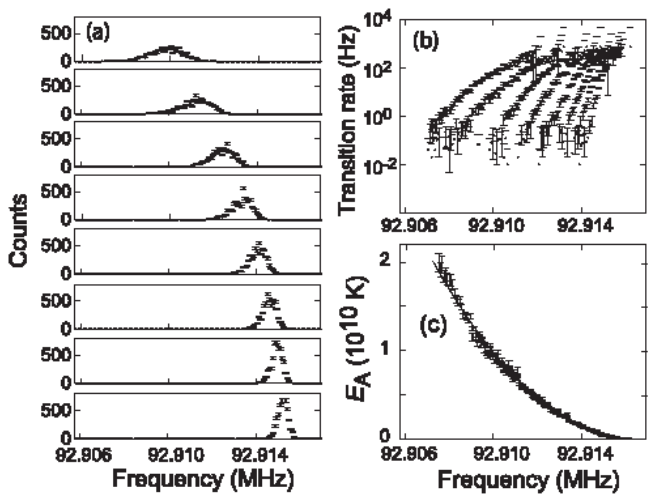} \caption{(a) Switching histograms
$h(\nu)$ for different noise powers, with ${\mathcal B} = $-56 dBm, 5 dB above the critical point,
for transitions from focus 1 to 2. (b) Transition rates $\Gamma(\nu)$ extracted from the switching
histograms. (c) Calculated energy barrier $E_A(\nu)$ extracted from the transition rates and the
variation in noise power. The noise power was varied from -127 dBm/Hz to -113 dBm/Hz. }
\label{fig:jitterhist}
\end{figure}

\begin{figure}
\includegraphics[width=1\linewidth]{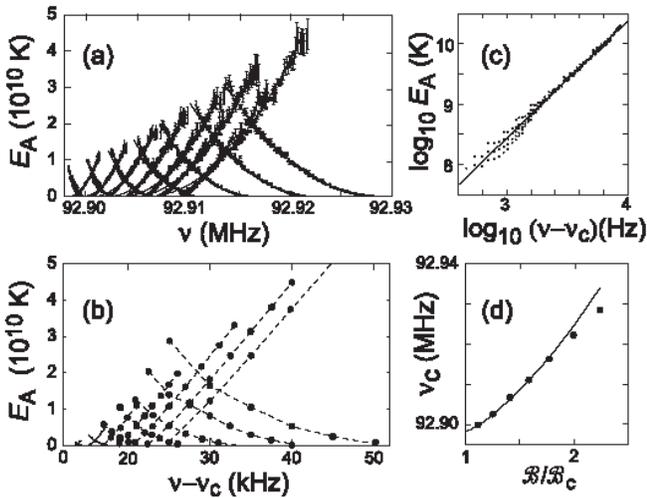}
\caption{(a) Measured quasienergy barrier for a drive force ranging from 1 dB to 7 dB above the
critical force ${\mathcal B}_c$. (b) Numerically-calculated energy barriers between foci 1 and 2,
assuming a Duffing oscillator with the same frequency, quality factor, and nonlinearity as measured
from Fig. \ref{fig:figure1}. (c) Log-log plot showing $E_A \propto (\nu-\nu_{c})^2$ dependence of
the energy barrier near the critical point. (d) $\nu_c$ versus drive amplitude ${\mathcal
B}/{\mathcal B})c$. At large amplitudes the data diverges from the analytic form.
\label{fig:answer}}
\end{figure}

In Fig. \ref{fig:jitterhist}(a) we display a set of histograms $h(\nu(t))$; higher noise powers
shift the peak switching frequency and also broaden the distribution. In Fig.
\ref{fig:jitterhist}(b) we show the transition rates extracted from these histograms, demonstrating
the rapid increase in transition rate as the noise power is increased. We then extract the
quasi-activation energy $E_A(\nu)$, by inverting the thermal activation expression $\Gamma(\nu)
\equiv \Gamma_{0} \exp (-E_A(\nu)/k_B T_{eff})$, where the effective temperature $T_{eff}$ is
proportional to the noise power, and the prefactor $\Gamma_{0}$ is related to the Kramers
low-dissipation form \cite{Kramers}, $\Gamma_{0} \approx \nu_0/Q$. We note that in this technique,
the histograms are only logarithmically sensitive to $\Gamma_{0}$, so that a precise determination
is difficult. In Fig. \ref{fig:jitterhist}(c) we display the activation energy $E_A(\nu)$ extracted
from the histograms, showing the expected decline in the barrier energy as the drive frequency
approaches the critical frequency. The distributions shown in Fig. \ref{fig:jitterhist}(b) are seen
to collapse onto a single curve $E_A(\nu)$. In Fig. \ref{fig:answer}(a) we show a collection of
experimentally measured $E_A(\nu)$ curves, measured for transitions from focus 1 to 2 and from 2 to
1, for different drive amplitudes.

We calculated the activation energies numerically. The dynamic solutions to Eq. (\ref{eq:duffing})
without noise give the relaxation from the separatrix to one of the foci. During a noise-induced
transition, the system is excited from a basin near a focus \emph{towards} the separatrix, which it
crosses and then relaxes to the other focus. There is an infinite number of possible trajectories
that allow a transition. Given a specific trajectory, it is possible to calculate the contribution
of the noise force using Eq. (\ref{eq:duffing}).  The total energy transferred to the resonator for
a particular trajectory is found by integrating the noise power along that trajectory, thus
yielding the effective quasienergy between the foci. The energy transferred is thus an action-like
quantity, and the most likely escape trajectory is that which requires the minimum action. The
action-like integral $S$ of the system is then $S = \int_{path} {\mathcal B}_{n}^2(t) dt$.

The most likely path $Y_0(t)$ minimizes the integral $S$. Because
the separatrix is a saddle point, the extremal trajectory will
most likely travel near the separatrix. The oscillator will
naturally evolve from a point near the separatrix to either focus,
without contributing to the action-like integral, as this
relaxation does not require a noise term. Only when the oscillator
is evolving against the dissipative flow field, from a focus
toward the separatrix, will it contribute to the action integral.

We used a numerical minimization of the possible trajectories $Y(t)$, using $S$ as a test function
to approach the extremum trajectory $Y_0(t)$.  Minimum trajectories were calculated for different
drive frequencies and amplitudes, yielding the energy barrier as a function of the drive amplitude,
shown in Fig. \ref{fig:answer}(b). We find good agreement (to logarithmic accuracy) between the
measured and calculated energy barriers.

Near the critical drive power ${\mathcal B}_c$, analytic forms indicate that the energy barrier
should have a quadratic dependence on the offset from the critical frequency $(\nu-\nu_{c})^2$
(where $\nu_c = \nu_{U,L}$) \cite{Dykman79} . This quadratic dependence is shown in Fig.
\ref{fig:jitterhist}(c) for one drive power, and for a range of different drive powers in Fig.
\ref{fig:answer}(c). This allows a determination of $\nu_c$ for a given drive power; our typical
histograms yield an uncertainty of $\Delta \nu_c/\nu_c \approx 3 \times 10^{-7}$. By comparing the
experimentally-observed dependence of $\nu_c$ on drive power with that obtained from numerical
analysis of the Duffing equation, shown in Fig. \ref{fig:answer}(d), we can extract the natural
resonance frequency $\nu_0$ and the coefficient of nonlinearity $K$; the former is the intercept of
the curves shown in that figure, and the latter related to the slope of the curves at small drive
powers. We find $\nu_0 = 92887360 \pm 10$ Hz and $K = (3745 \pm 4) \times 10^{11}$ N/m$^3$. The
frequency measurement represents a relative precision of $\Delta \nu_0/\nu_0 \approx 1.1 \times
10^{-7}$. This level of frequency resolution has significant implications for e.g. mass sensing
with mechanical resonators \cite{Ekinci:2004b, Ilic:2004}.

\begin{figure}
\hspace{.3in}
\includegraphics[width=\linewidth]{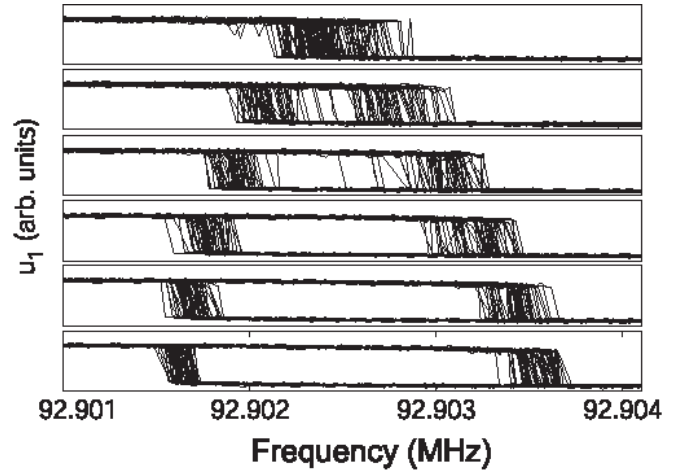}
\caption{Amplitude hysteresis plots, for no noise power (bottom), with the drive amplitude set at
-59 dBm, 2 dB above the critical point. The noise power was increased in 2 dB steps for each
succeeding frame.  At the largest noise power, the hysteresis is quenched.} \label{fig:hysteresis}
\end{figure}

The measurements described above were made in the small-to-moderate noise limit, with noise
energies much less than the energy barrier. At higher noise powers, the hysteresis due to the
nonlinear response can actually be quenched, by rapid noise-induced transitions between the two
foci. This quenching is demonstrated in Fig. \ref{fig:hysteresis}: As the noise power is increased,
the area of the hysteresis loop grows visibly smaller, until, at the highest noise powers, the
switching is no longer hysteretic. In this limit, the oscillator generates random telegraph signals
as it makes transitions from one focus to the other. The spectrum of the random telegraph signal is
related to the transition rate of the oscillator.

In conclusion, we have measured the configuration space trajectories, and the transition rates,
between the bistable states of a nonlinear radiofrequency mechanical resonator. These measurements
are in good agreement with numerical simulations based on the Duffing oscillator equation of
motion. Detailed analysis of the noise-induced switching transitions allows a quantitative
measurements of the energy barrier between the stable foci, and provides a highly sensitive
measurement of two key resonator parameters, the resonator natural frequency and the nonlinear
parameter.

\textbf{Acknowledgments.} We would like to thank C.S. Yung, R. G. Knobel, D. R. Schmidt, L. J.
Swenson, and D. K. Wood for their support. This work was funded by the DARPA-DMEA/UCSB Center for
Nanoscale Innovation for Defense.

\clearpage
\bibliography{paperbib}

\end{document}